\documentclass[floatfix,10pt,onecolumn,showpacs,preprintnumbers,amsmath,amssymb]{revtex4}

\usepackage{epsf}
\usepackage{graphicx}  % Include figure files
\usepackage{dcolumn}   % Align table columns on decimal point
\usepackage{bm}        % bold math

\newcommand{\be}{\begin{equation}}
\newcommand{\en}{\end{equation}}
 \newcommand{\bea}{\begin{eqnarray}}
 \newcommand{\ena}{\end{eqnarray}}
  \newcommand{\sch}{Schwarzschild}

\begin{document}

\title{From thermodynamics to the solutions in gravity theory}
\author{Hongsheng Zhang$^{1,2~}$\footnote{Electronic address: hongsheng@shnu.edu.cn},  Xin-Zhou Li$^1$ \footnote{Electronic address: kychz@shnu.edu.cn} }
\affiliation{ $^1$Center for
Astrophysics, Shanghai Normal University, 100 Guilin Road,
Shanghai 200234, China\\
$^2$State Key Laboratory of Theoretical Physics, Institute of Theoretical Physics, Chinese Academy of Sciences, Beijing, 100190, China
}

\date{ \today}

\begin{abstract}
 In a recent work, we present a new point of view to the relation of gravity and thermodynamics, in which we derive the \sch~solution through thermodynamic laws by the aid of the Misner-Sharp mass in an adiabatic system. In this paper we continue to investigate the relation between gravity and thermodynamics for obtaining solutions via thermodynamics. We generalize our studies on gravi-thermodynamics in Einstein gravity to modified gravity theories. By using the first law with the assumption that the Misner-Sharp mass is the mass for an adiabatic system, we reproduce the Boulware-Deser-Cai solution in Guass-Bonnet gravity. Using this gravi-thermodynamics thought, we obtain a NEW class of solution in $F(R)$ gravity in an $n$-dimensional (n$\geq$3) spacetime which permits three-type  $(n-2)$-dimensional maximally symmetric subspace, as an extension of our recent three-dimensional black hole solution, and four-dimensional Clifton-Barrow solution in $F(R)$ gravity.

\end{abstract}


%%\preprint{arXiv: }
 \maketitle

\section{Introduction}

    Gravity is inherently related to  thermodynamics. The deep and extensive relations have been explored more for 40 years. Gravi-thermodynamics originates from the studies of black holes. In 1970's, a series of important discoveries implies that black holes is in fact a thermodynamic object, through they are controlled by gravity, for more original references, see \cite{wald}.

    Inspired by black hole thermodynamics, thermodynamic laws is shown to valid in several other systems commanded by gravity, such as wormhole, the universe etc. The successes of the applications of thermodynamics to these objects in gravitational theory bring confidence to us that gravity theory itself may contain the information of thermodynamic theory, more or less. However, a general demonstration of this point is difficult since even the physical quantities like mass, entropy, and angular momentum do not make sense in a generic spacetime described by the Einstein field equation. No conserved charge is well-defined if there is no Killing field in the spacetime \footnote{In some special symmetric situations, for example in a spherically symmetric spacetime, we can define a conserved charge by using the Kodama vector. But a conserved charge is always associated with a Killing vector in a general spacetime.}. It seems impossible to make a general discussion about the relation between gravity and thermodynamics.  An interesting idea in this direction is to consider the inverse problem: Does thermodynamics imply gravity theory?

     Jacobson made the first try to obtain gravity theory from thermodynamics \cite{jaco}.  He  derives Einstein equation on a hypersurface tilting to a null surface by using the local first law of equilibrium thermodynamics.  The entropy is assumed to be proportional to the area of the local Rindler horizon of an infinitely accelerated Rindler observer.  The temperature of  the system in consideration is taken as the Unruh temperature sensed by this observer. This study starts from equilibrium thermodynamics, but the resulted Einstein equation can be used in a general case. One  is justified to guess that there is some technical problem in the reasoning. The technical details of this problem are pointed out in \cite{ted2}.  The order of the local Killing vector is displayed to be problematic. Verlinde suggest an entropic force method, which can also derive Einstein field equation from thermodynamics \cite{ver}. Different from Jacobson' approach, Verlinde supposes a stationary spacetime from the very beginning. It seems reasonable to assume the spacetime to be in equilibrium, and thus the application of equilibrium thermodynamics is guarantied.  The entropic force approach has been extended to several other cases \cite{several}.  However, some problems have been found in the entropic force method. For instance, the experiments by using ground based ultra-cold neutrons seem contradicting to the concept of  entropic force \cite{av}.

    We make a new observation on the relation between gravity and thermodynamics in a recent work \cite{self}, in which for the first time we derive the \sch~ solution directly from thermodynamic laws without invoking Einstein field equation. In the demonstration, we do not borrow any concept from quantum theory. In this paper, we extend our previous studies to the cases of modified gravity theories.

    This paper is organized as follows. In the next section we revisit the main results of the previous work. Also, we give some closely related new results. We shall show that the other two topological cases of \sch~ solution can be obtained via almost the same reasoning. We find a new way to get de Sitter/anti de Sitter solution without including the density and pressure of the matter field.  In section 3, we derive the Boulware-Deser-Cai solution in Gauss-Bonnet gravity. In section 4, we derive a class of solution in $F(R)$ gravity in an $n$-dimensional (n$\geq$3) spacetime which permits three-type  $(n-2)$-dimensional maximally symmetric subspace. This is a really new one. When $n=4$, it degenerates to the static Clifton-Barrow solution in $F(R)$ theory. When $n=3$, it degenerates to a special case of our recent black hole solution \cite{self2}. We conclude this paper in section 5.

    \section{revisit the derivation of \sch~}

   First, we consider a static space which permits a two-dimensional maximally symmetric subspace with 3 types of sectional curvatures $k=1,~0,~-1$,
   \be
   ds^2=-f(r)dt^2+h(r)dr^2+r^2d\Omega_2^2,
   \label{2dim}
   \en
    in which $f(r),~h(r)$ are general functions of $r$,  $\Omega_2$ denotes a unit two-sphere, two-cube, or two-pseudo-sphere, depending on the sectional curvature $k=1,~0,~-1$, respectively. Our previous work only deals with the case of $k=1$. Here we consider the 3 cases at the same time. The Misner-Sharp mass form used in \cite{self} to derive \sch~ and related solutions is the form for Einstein gravity without a cosmological constant in a spherically symmetric space.

   Misner and Sharp proposed their mass form in the studies of gravitational collapse \cite{ms}. The  Misner-Sharp mass permitting different topologies (without cosmological constant) reads \cite{cai3},
   \be
    M_{ms}=\frac{1}{2G }(k-I^{ab}\partial _{a}r\partial
_{b}r),
  \label{MSo}
     \en
     where
     \be
     I=-f(r)dt^2+h(r)dr^2.
     \en
  Considering a four-dimensional adiabatic Misner-Sharp system,
  we write the first law as,
   \be
   \delta M_{ms}=0.
   \label{ad}
   \en
   Thus we obtain,
   \be
   k-h-rh'=0,
   \en
    where a prime denotes the derivative with respect to $r$.
    Then immediately we reach,
    \be
    h=k-\frac{C}{r},
    \label{h}
    \en
    where $C$ is an integration constant. Substituting $h$ into (\ref{MS4}), we find the physical sense of $C$,
    \be
    C=2M_{ms}.
    \en
    In this way, we obtain $h$  with 3 kinds of sectional curvatures as a unity.

      Next, we use surface gravity to determine $f(r)$. As usual, we calculate the surface gravity $\kappa$ by the product of $(-g_{00})^{1/2}$ and the magnitude of the
       4-acceleration for a particle resting at the static coordinates. The result is,
     \be
     \kappa=\frac{1}{2}(fh)^{-1/2}f'.
     \label{kappa}
     \en
    For a detailed demonstration, see \cite{self}.

    In 4 dimensional spacetime, the geometric surface gravity corresponding to the unified first law reads,
     \be
     \kappa=\frac{M_{ms}}{r^2}-4\pi rw.
     \label{kapp}
     \en
      Here $w$ is the work term,
     \be
     w=-\frac{1}{2}I^{ab}T_{ab}.
     \en
        $T$ is the stress energy of the matter fields. For vacuum case, $T=0$. We note that the definition of ($\ref{kapp}$) is independent on the sectional curvature of $\cal K$.
      From (\ref{kappa}) and (\ref{kapp}), we have
       \be
       \frac{1}{2}(fh)^{-1/2}f'=\frac{M_{ms}}{r^2}.
       \label{sureq}
       \en
       Then we arrive at
       \be
       f=\left[\left(k-\frac{2M_{ms}}{r}\right)^{1/2}+D\right]^2,
       \en
       by using (\ref{h}), $D$ is an integration constant. Similar to the previous work, we find $D=0$ according to the Newtonian condition or Minkowskian condition.

         We see that the derivation topological ~\sch~ spacetime via thermodynamics is almost the same as that of \sch~case. The only difference is the starting metric and the corresponding Misner-Sharp mass take different forms, as shown in (\ref{2dim}) and (\ref{MSo}).

       Then, we revisit asymptotical (anti) de Sitter spaces. De Sitter or anti de Sitter depends on the sign of the cosmological constant. In this case, we make a different thought on the matter field. There are always two perspectives on the cosmological constant. The first one is that it is just the  vacuum, whose stress energy satisfies that energy density equals negative pressure. The Misner-Sharp mass is taken to be the original form, i. e., the form without a cosmological constant. The second one is that the cosmological constant belongs to the gravitational field rather than the matter fields. However, in this perspective, we must use a Misner-Sharp mass form with cosmological constant. We have demonstrated how to reach \sch-(anti) de Sitter metric in the first perspective in the previous work \cite{self}. Now we try to derive it in the second perspective. First we present the Misner-Sharp mass from with a cosmological constant.
      In an $n$-dimensional spacetime $\cal M$=$\cal I \times K$,
    \be
  ds^2=-f(r)dt^2+\frac{1}{h(r)}dr^2+r^{2}\gamma _{ij}dz^{i}dz^{j},
  \label{metrics}
  \en
    where (${\cal K},~\gamma$) is an $(n-2)$-dimensional maximally symmetric submanifold embedded in (${\cal M}, g$).

     The Misner-Sharp mass for Einstein gravity with a cosmological constant for the above metric within radius $r$ reads \cite{self3},
    \be
       M_{ms}=\frac{V_{n-2}^{k}r^{n-3}}{8\pi G }\left[\frac{n-2}{2}(k-I^{ab}\partial _{a}r\partial
_{b}r)-\frac{r^2\Lambda}{n-1}\right],
  \label{MSs}
  \en
  where $V_{n-2}$ is the volume of a unit submanifold $\cal K$, $I$ is the induced metric on $\cal I$, and the indexes $a,~b$ run from 0 to 1. When $n=4$, it reduces
  to
   \be
   M_{ms}=\frac{V_{2}^{k}r}{8\pi G }\left[(k-I^{ab}\partial _{a}r\partial
_{b}r)-\frac{\Lambda r^2}{3}\right].
  \label{MS4}
  \en
  A system is called Misner-Sharp system, if its gravitational energy is defined as its Misner-Sharp mass.
       Assuming the Misner-Sharp system is in adiabatic state, and making a variation with respective to $r$ in (\ref{MS4}) with a cosmological constant, we obtain,
      \be
       k-h-\frac{\Lambda}{3}r^2=r\left(h'+\frac{2}{3}\Lambda r\right).
       \en
      It is fairly easy to get the solution,
      \be
      h=k-\frac{C}{r}-\frac{\Lambda}{3}r^2.
      \label{hde}
      \en
       To obtain the Misner-Sharp mass in the asymptotical (anti) de Sitter space, one substitutes (\ref{hde}) into (\ref{MS4}),
      \be
      M_{ms}=\frac{C}{2},
      \en
      which is different from the result in the first perspective, in which
      \be
      \overline{M}_{ms}=\frac{C}{2}+\frac{\Lambda}{6}r^3.
      \en
      The physical interpretation of the difference is that $\Lambda$ is the gravitational back ground and has been subtracted in (\ref{MS4}) in the first perspective, but in the second perspective $\Lambda$ is treated as the
      matter field, which has been included in the Misner-Sharp mass.

     Following the previous investigations, the next step should be to compare the surface gravity (\ref{kappa}) with the definition of surface gravity in the unified first law (\ref{kapp}) to derive $f$. But, unfortunately we do not have a proper definition of surface gravity in modified gravity as an extension of (\ref{kapp}). Here, a theory permitting any deviation from Einstein theory (no cosmological constant) is treated as modified gravity, since the original Misner-Sharp mass used in the unified first law does not permit a cosmological constant \cite{hayward1} \cite{hayward2}. In this sense, a gravity with a cosmological constant is regarded as a ``modified gravity". So we have to switch to the first perspective, i.e., to treat the cosmological constant as a vacuum matter with density and pressure,
     \be
     \rho=\frac{\Lambda}{8\pi G},
     \en
     \be
     p=-\frac{\Lambda}{8\pi G}.
     \en
     The following procedure exactly mimics what we done in \cite{self}. The resulted $f$ is
      \be
      f=k-\frac{C}{r}-\frac{\Lambda}{3}r^2.
      \label{fde}
      \en
     It seems that we take a roundabout way and make superfluous arguments. In fact the switching between two perspectives is critical in search for solution in modified gravities via thermodynamic considerations. We shall see its power in the following sections.
   \section{Gauss-Bonnet gravity}
   In four-dimensional spacetimes,  the only form of Lagrangian which generates field equations in absence of higher than two order derivatives with respect to metric is the Hilbert-Einstein action together with a cosmological constant. But in higher dimensional spacetime, one finds a proper combination of $R^2$-type which also does not yield higher than two order derivatives with respect to metric \cite{love}, which is called Gauss-Bonnet term $R_{GB}$
   \be
      R_{GB}=R_{\mu\nu\gamma\delta}R^{\mu\nu\gamma\delta} -4 R_{\mu\nu}
   R^{\mu\nu}+R^2,
      \en
   where $R_{\mu\nu\gamma\delta},~R_{\mu\nu},$ and $R$ denote Reimann tensor, Ricci tensor, and Ricci scalar, respectively. In four or lower dimensional spacetime, Gauss-Bonnet is a surface term, which does not appear in the resulted field equation. More general Lovelock gravity permits combinations of higher order power of $R$, such as $R^4$-type terms, which can also evade higher than two order derivatives with respect to the metric \cite{love}. In type II-B and heterotic string theories, the Gauss-Bonnet term is the next to leading order correction to Hilbert-Einstein term. The Gauss-Bonnet term is free of ghosts when expanding on the Minkowskian background, without the problems of unitarity. A static, spherically symmetric solution of Gauss-Bonnet gravity is obtained in \cite{deser}. This solution is extended by including a cosmological constant \cite{caigb}. We call this solution  Boulware-Deser-Cai solution. Here we use thermodynamic method to derive this solution.

   Our starting point is the metric (\ref{metrics}) on $\cal M$. The Misner-Sharp mass for this metric is given in \cite{mae}, see also \cite{caims},
   \be
   M_{ms}=\frac{V_{n-2}^{k}r^{n-3}}{8\pi G }\left[\frac{n-2}{2}(k-I^{ab}\partial_{a}r\partial_{b}r)-
{\frac{\Lambda }{n-1}}r^{2}+\frac{n-2}{2}{\tilde {\alpha
}}r^{-2}(k-I^{ab}\partial_{a}r\partial_{b}r)^{2}\right],
  \label{msgb}
  \en
  which corresponds to the action,
  \be
  S=\frac{1}{16\pi G}\int_{\cal M} d^n x\sqrt{-\det (g)}\left(R -2 \Lambda
  + \alpha R_{GB}\right),
   \en
 in which
 \be
  \tilde {\alpha}=\alpha (n-3)(n-4).
  \en
 The reductions from other theories, such as string theory, to the Gauss-Bonnet gravity may impose some constraints on $\alpha$. Here, we just treat is as a free parameter.
 Considering an adiabatic Misner-Sharp system, we have
 \be
 \delta M_{ms}=0,
 \en
  which yields from (\ref{msgb}),
  \be
  \frac{n-3}{r} \left(k-\frac{2 r^2 \Lambda }{2-3 n+n^2}+\frac{\tilde{\alpha}  (k-h)^2}{r^2}-h\right)=\frac{4 r \Lambda }{2-3 n+n^2}+\frac{2 \tilde{\alpha} (k-h)^2}{r^3}+h'+\frac{2 \tilde{\alpha}  (k-h) h'}{r^2},
  \en
  where a prime denotes derivative with respect to $r$. On the face of it, this equation seems complicated. But it is in fact only a first-order equation. Direct integration presents,
  \be
 h= k +\frac{r^2}{2\tilde\alpha}\left ( 1 \mp
 \sqrt{1+\frac{8\tilde\alpha\Lambda}{(n-2)(n-1)}+\frac{4{\tilde\alpha}^2 C}{r^{d-1}}
   } ~\right),
   \label{meGB}
   \en
   where $C$ is an integration constant. Back substituting (\ref{meGB}) into (\ref{msgb}), one clears the physical sense of $C$,
   \be
   M_{ms}= \frac{\tilde\alpha C (n-2)V_{n-2}^{k}}{16\pi G }.
   \en
  At the limit $\alpha\to 0$ and $\Lambda\to 0$, one can confirm $M_{ms}$ is exactly the mass in Newtonian sense. So the Misner-Sharp mass seems a reasonable generalization of Newtonian mass in Guass-Bonnet gravity.

  Next, we try to work out $f$. We confront the same problem as that in the case of cosmological constant: there is no proper definition of surface gravity in modified gravity. Similar to the case of the cosmological constant, we also have two perspectives about modified gravity theory. The first perspective is to treat all the terms generated by the action other than Einstein tensor as stress energy of ``matter fields". Actually, this perspective has been extensively explored in cosmology. Usually, all the terms other than Einstein tensor is treated as effective dark energy to investigate the cosmic acceleration \cite{selfcos}, where this perspective is called Einstein interpretation. The second perspective is to treat all the geometric sector as gravity. That is the natural modified gravity perspective.

    We have derived $h$ in the second perspective. Now we switch to the first perspective (Einstein interpretation). The effective stress energy $T^{(e)}$ in the first perspective is \cite{caigb},
    \be
    T^{(e)}_{\mu\nu}=-\frac{1}{8\pi G}\left[2\alpha (RR_{\mu \nu }-2R_{\mu \alpha }R_{\nu }^{\ \alpha
}-2R^{\alpha \beta }R_{\mu \alpha \nu \beta } +R_{\mu }^{\ \alpha
\beta\gamma }R_{\nu \alpha \beta \gamma
})-\frac{\alpha}{2} g_{\mu\nu}R_{GB}+\Lambda g_{\mu \nu }\right],
   \label{effT}
    \en
      where the curvature tensors correspond to the metric (\ref{metrics}). In this perspective, the Misner-Sharp mass takes its original form,
      \be
       \overline{M}_{ms}=\frac{V_{n-2}^{k}r^{n-3}}{8\pi G }\frac{n-2}{2}(k-I^{ab}\partial _{a}r\partial
_{b}r),
  \label{MSso}
  \en
  which is no more to be a constant, since the ``matter field" (\ref{effT}) is included. Similar to the case of real matter field, the work term is defined as
  \be
  w=-\frac{1}{2}I^{ab}T^e_{ab}.
  \label{worke}
  \en
  With these preparations, we define the surface gravity in $n$-dimensional ($n\geq 3$) spacetime
  \be
  \kappa\equiv \frac{8\pi (n-3)}{(n-2)V_{n-2}^{k}}\frac{\overline{M}_{ms}}{r^{n-2}}-\frac{8\pi}{n-2}rw.
  \en
  One is easy to check that when $n=4$, it degenerates to (\ref{kapp}). Almost throughout this section we suppose $n\geq 5$ since we are discussing the Gauss-Bonnet gravity. We would like to point out that the extent of application of the above equation is $n\geq 3$. The case with $n=3$ has special significance.  When $n=3$, $\kappa$ vanishes for pure Einstein gravity, which is consistent with the important result in three-dimensional gravity: There is no black hole with nontrivial geometries for three-dimensional Einstein gravity. However, when work term appears, the black holes with non-trivial geometries are also possible in three-dimensional gravity.

  Substituting (\ref{MSso}) and (\ref{worke}) into the above equation, we reach,
  \be
  \kappa=\frac{1}{2G}(n-3)\frac{1-h}{r}+\frac{4\pi r}{n-2}I^{ab}T^e_{ab}~.
  \label{kappanew}
  \en
  From (\ref{kappa}), and (\ref{kappanew}), we obtain the equation for $f$,
  \be
  \left(1-2 \sqrt{\frac{f}{h}}~\right) h f'+fh'=0.
  \en
 Then substituting the expression for $h$ in (\ref{meGB}), we obtain $f$,
 \be
 f=\frac{1}{2} \left(h-D\pm \sqrt{h^2-2Dh}~\right),
 \en
 where $D$ is an integration constant. To determine $D$, we explore some limits of $f$.
 At the limit $\Lambda \to 0$ and $\alpha\to 0$, it should degenerate to \sch~solution, which we have find by thermodynamics. It is easy to check $D=0$ in the ``$+$" branch gets correct \sch~limit. The ``$-$" branch is an extraneous solution generated by this thermodynamic method. Thus we find $f$,
 \be
 f=h= k +\frac{r^2}{2\tilde\alpha}\left ( 1 \mp
 \sqrt{1+\frac{8\tilde\alpha\Lambda}{(n-2)(n-1)}+\frac{4{\tilde\alpha}^2 C}{r^{d-1}}
   } ~\right).
   \en
 Thus we complete the Boulware-Deser-Cai solution in Gauss-Bonnet gravity based on thermodynamic considerations. In principle, it opens a new window to explore the inherent relations between gravity and thermodynamics. In practice, one sees that all the equations we should solve are first-order equations, which is easier than to solve the field equations directly.

 \section{$F(R)$ gravity}

 Gauss-Bonnet gravity, and more general Lovelock gravity, do not generate higher than two order derivatives with respect to metric in the field equation, though they contain $R^2$ or more higher order terms. Divergence is an old and hard problem in gravity. It is found that the divergences are drastically alleviated if higher order derivatives are introduced \cite{stelle}.
 One will also meet such terms when one considers quantum effects \cite{bir} or reduced gravity from other theory, for example string theory \cite{string}. $F(R)$ gravity is one of the most extensively studied theory in the  theories with higher order derivatives. $F(R)$ gravity has some distinctive properties. First of all, it is the unique one which successfully extricates from the catactrophic Ostrogradski instability amongst all higher derivative gravity theories \cite{wood}. Second, it is simple enough to handle, at the same time complicated enough to support the principle framework of higher derivative theories.

 We shall derive a new solution for $F(R)$ theory via thermodynamics. The Misner-Sharp mass for $F(R)$ gravity in 4 dimensional spherically symmetric spacetime is presented in \cite{cai3}. We obtain the general form of Misner-Sharp mass in $F(R)$ gravity  in an $n$-dimensional spacetime  with 3 types of $(n-2)$-dimensional maximally symmetric submanifold \cite{self3}. Our starting point is still a Misner-Sharp system. The metric is given in (\ref{metrics}). We work on this $n$-dimensional manifold (${\cal M}, g$) with an $(n-2)$-dimensional maximally symmetric submanifold (${\cal K},~\gamma$), on which the Misner-Sharp mass reads,
  \bea
  M_{ms}&=&    \frac{V_{n-2}^{k}r^{n-3}}{8\pi G }\Big[\frac{n-2}{2}(k-I^{ab}\partial _{a}r\partial_{b}r)F_R+\frac{1}{2(n-1)}%
r^{2}(F-F_RR)- rI^{ab}\partial _{a}F_R\partial _{b}r\Big]~\notag \\
&+&\frac{V_{n-2}^{k}}{16\pi G }\int dr\Big[r^{n-2}h'+(n-2)r^{n-3}(h-k)+\frac{r^{n-1}}{n-1}R\Big]F_{R,r},
\label{MSs}
  \ena
 which corresponds to the action,
 \be
   S=\frac{1}{16\pi G }\int_{\cal M} d^{n}x\sqrt{-{\rm det}(g)}~F(R)+S_{m},
  \label{actionfR}
  \en
 where $S_{m}$ is the action of the matter fields, and $F_R=\partial F(R)/\partial R$. Considering the vacuum case $S_m=0$, in which the Misner-Sharp system is adiabatic, we have
   \be
   \delta M_{ms}=0,
   \en
   which yields,
   \bea
   r^2 d R^3 &+& (1+d) (n-2) R^2 \left[3-n+(n-3) h+r h'\right]+2 d (d^2-1) r^2 h R'^2\\
   &+& d (1+d) r R \left[r h' R'+2 h \left((n-2) R'+r R''\right)\right]=0,
   \label{hrfr}
   \ena
   where we take $F(R)=R^{d+1}$~, and the Ricci scalar is given by
   \be
    R=\frac{(n-3)(n-2)}{r^2}(k-h)-\frac{n-2}{r}\left(h'+\frac{hf'}{f}\right)+\frac{1}{2f}\left(\frac{hf'^2}{f}-h'f'-2hf''\right).
   \label{RS}
   \en
   One sees that different from the cases of Einstein gravity and Gauss-Bonnet gravity, (\ref{hrfr}) is not an equation of a single function, since $f(r)$ enters the equation through $R$. Furthermore, it is a high order equation after substituting $R$ into (\ref{hrfr}). Principally, we can use the first perspective (Einstein interpretation) to obtain the equation of $\kappa$. And we then use the associated equations of (\ref{hrfr}) and $\kappa$ to find the two functions $h(r)$ and $f(r)$. However, it is hard to find the analytical solutions in this way.

   Observing (\ref{RS}) carefully, and from the experiences three-dimensional black hole \cite{self2} and four-dimensional black hole \cite{CB} in $F(R)$ gravity, we make a tentative ansatz,
  \be
  R=-\frac{kL}{r^2},
  \label{conje}
  \en
  where $L$ is a constant.
 Using this ansatz, (\ref{hrfr}) becomes tractable. The solution is
 \be
 h=\frac{6-5 n+n^2+d \left(6+L-5 n+n^2\right)}{(1+d) \left(6+8 d^2-4 d (n-3)-5 n+n^2\right)}\left(k+Cr^{3+2 d+\frac{2 d (1+2 d)}{2+2 d-n}-n}\right).
    \label{hsolu}
    \en

  Then we switch to the first perspective. The effective stress energy reads \cite{soti},
  \be
  T^e_{\mu\nu}=\frac{1}{8\pi G F_R}\Big[%
\frac{1}{2}~g_{\mu \nu }(F-RF_{R})+\nabla _{\mu }\nabla _{\nu
}F_R-g_{\mu \nu }\square F_R\Big].
   \label{tefr}
     \en
  The $\kappa$ in (\ref{kappanew}) becomes really involved.  From (\ref{RS}) and (\ref{conje}), we have,
   \be
  \frac{(n-3)(n-2)}{r^2}(k-h)-\frac{n-2}{r}\left(h'+\frac{hf'}{f}\right)+\frac{1}{2f}\left(\frac{hf'^2}{f}-h'f'-2hf''\right)=-\frac{kL}{r^2}.
  \en
  Substituting $h$ in (\ref{hsolu}), we obtain $f$,
  \be
  f= r^{\frac{2 d (1+2 d)}{1+d-n/2}}\left(k+C r^{\frac{6+8 d^2-4 d (n-3)-5 n+n^2}{2+2 d-n}}\right),
  \en
  and rewrite $h$
  \be
  h=\frac{(3-n) (n-2-2 d)^2}{\left(2+4 d+4 d^2-n\right) \left(6+8 d^2-4 d (n-3)-5 n+n^2\right)} \left(k+C r^{\frac{6+8 d^2-4 d (n-3)-5 n+n^2}{2+2 d-n}}\right),
    \en
  in which all the integration constants and $L$ have been calibrated by Clifton-Barrow solution,
     \be
   L=-\frac{(n-3)(n-2)d(d+1)}{1/2-n/4+d(d+1)}.
  \en
   One can check that this solution degenerates to the Clifton-Barrow solution when $n=4$. We can confirm that it satisfies the vacuum field equation of $F(R)$ gravity \cite{self5},
      \be
F_RR_{\mu \nu }-\frac{1}{2}Fg_{\mu \nu }-\nabla _{\mu }\nabla
_{\nu }F_R+g_{\mu \nu }\square F_R=0.
\label{fieldfr}
\en
   The physics of higher dimensional case of this solution may be of interests. Here we first say something about the three-dimensional case of this solution. At first sight, one may think that this solution is trivial since both $h$ and $R$ vanish. However, a special case with
   \be
   d=\frac{1}{2} \left(-1\pm\sqrt{2}~\right),
   \en
 is none trivial. In the three-dimensional case, the submanifold (${\cal K},~\gamma$) is an one-dimensional cube, i.e., a line. Thus in principle, the three cases of (${\cal K},~\gamma$) merge in a local geometric view. Globally, the topologies of the whole manifolds (${\cal M},~g$) permitting different submanifolds with different $k$ are different.  Under this condition, $f$ and $h$ read,
  \be
  f=kr^2+C r^{\sqrt{2}},
    \label{f1}
    \en
  \be
  h=k+C r^{-2+\sqrt{2}}.
  \label{h1}
  \en
  This a real black hole with true singularity, which is completely different from the case of three-dimensional Einstein gravity, where it is no real black hole with non trivial geometries. A more general three-dimensional black hole solution in $F(R)$ gravity has been derived in our recent work \cite{self2}, where the the corresponding components of the metric read,
  \be
  f=hr^{\frac{2pd(1+pd)}{1-pd}},
  \label{Ar}
  \en
  \be
   h=k\frac{L(1-d  p) r^{2-p}}{2 (d +1) \left(d ^2 (2 d +1) p^3-p+2\right)} + C r^{\frac{2 p^2 d ^2}{p d -1}},
   \label{Br}
   \en
   and
  \be
  p=\frac{1-2d}{(1+2d)d},
  \en
  which describes the first three-dimensional vacuum black hole with non-trivial geometry. It is easy to check that the metric components in (\ref{f1}) and (\ref{h1}) are a special case of (\ref{Ar}) and (\ref{Br}) with $p=2$ ($d=\frac{1}{2} \left(-1\pm\sqrt{2}~\right)$).

   \section{conclusion}
    The relation between gravity and thermodynamics has been a research focus in physical society for 40 years. In a recent work, we present a new view on this relation. We can derive \sch~solution from thermodynamic considerations \cite{self}. There are two key points in our demonstrations: The first one is an adiabatic Misner-Sharp system, and the second one is the surface gravity defined according to the unified first law. In this paper, we generalize this investigation to the case of modified gravities, and obtain some new results in modified gravity.

    We find that topological \sch~ solution can be derived via almost the same considerations. In the asymptotic de Sitter/anti de Sitter case, we show that the component $h$ also can be obtained in the second perspective (the modified gravity perspective). In the Gauss-Bonnet gravity, we derive the Boulware-Deser-Cai solution using a similar considerations. We first get $h$ in the second perspective in an adiabatic Misner-Sharp system. And then we switch to the first perspective to obtain $f$ by using an equality of surface gravity.
    Recently, $F(R)$ gravity gets more and more attentions. Its foundation  and applications in cosmology have been extensively studied. We present the Misner-Sharp mass in arbitrary dimension ($n\geq 3$) for $F(R)$ gravity. Using this form, we successfully obtain a NEW class of solution for $R^{d+1}$ gravity. When $n=4$, it reduces to the Clifton-Barrow solution.
    For a special $d$, $4d(d+1)=1$, the three-dimensional solution reduces to a special case of a more general black hole in our previous work \cite{self2}.

    In principle, this study opens a new window to explore the relation between gravity and thermodynamics. The quasilocal mass form, especially the Misner-Sharp mass, may hide rich information of the gravity field. In practise, it offers a new method to solve the field equation. As we have seen in the previous sections, the equations appeared in this thermodynamic demonstration are usually first order equation. They may be easier than to solve the field equation directly.

 {\bf Acknowledgments.}
   This work is supported by the Program for Professor of Special Appointment (Eastern Scholar) at Shanghai Institutions of Higher Learning, National Education Foundation of China under grant No. 200931271104, and National Natural Science Foundation of China under Grant No. 11075106 and 11275128.

\end{document}